\documentclass[cameraready]{Interspeech}

\usepackage{comment}
\usepackage{multirow}
\usepackage{subcaption}

\title{GLAD: Global-Local Aware Dynamic Mixture-of-Experts for Multi-Talker ASR}

\author[affiliation={1}, orcid=0009-0003-0872-4249]{Yujie}{Guo}
\author[affiliation={1}, orcid=0009-0002-4819-4572]{Jiaming}{Zhou}
\author[affiliation={1}, orcid=0009-0001-2407-0789]{Yuhang}{Jia}
\author[affiliation={1}, orcid=0000-0001-5068-025X]{Shiwan}{Zhao}
\author[affiliation={1}, orcid=0009-0000-2748-3020, correspondingauthor]{Yong}{Qin}

\address{
    $^1$ TMCC, College of Computer Science, Nankai University, Tianjin, China 
}

\email{guoyujie02@mail.nankai.edu.cn, qinyong@nankai.edu.cn}

\keywords{multi-talker ASR, mixture of experts, cocktail party problem, dynamic routing}

\begin{document}

\maketitle

\begin{abstract}
End-to-end multi-talker automatic speech recognition (MTASR) faces significant challenges in accurately transcribing overlapping speech. A critical bottleneck is that speaker-specific acoustic characteristics, which are essential for distinguishing overlapping speech, are often diluted in deep network layers. To address this, we propose the Global-Local Aware Dynamic Mixture-of-Experts (GLAD) architecture. GLAD introduces a novel routing mechanism that dynamically fuses speaker-aware global context with fine-grained local acoustic details to adaptively guide expert selection. Experiments on the LibriSpeechMix and CH109 datasets demonstrate that GLAD significantly outperforms existing Serialized Output Training (SOT)-based MTASR approaches, exhibiting exceptional robustness in challenging, high-overlap scenarios. To the best of our knowledge, this is the first work to apply a global-local fusion MoE strategy to MTASR.

\end{abstract}

\section{Introduction}
\label{sec:Introduction}
Multi-talker automatic speech recognition (MTASR) represents a critical challenge in speech processing, requiring models to simultaneously decode overlapping utterances from multiple speakers in complex acoustic environments. This capability is essential for practical applications including meeting transcription and multi-party dialogue analysis, where speakers frequently interrupt or talk simultaneously.

Current approaches to MTASR can be broadly categorized into two paradigms. Single-input multiple-output (SIMO) methods~\cite{PIT3, PIT5-trans, PIT6} explicitly separate mixed speech into speaker-specific branches before transcription, typically employing permutation invariant training (PIT)~\cite{PIT1, PIT2}, which incurs additional computational cost. To improve efficiency, heuristic error assignment training (HEAT)~\cite{HEAT1, HEAT2} applies the Hungarian algorithm for optimal permutation assignment. However, SIMO methods assume a fixed number of speakers and require explicit separation, limiting their applicability in real-world scenarios with unknown or variable speaker counts.

Alternatively, single-input single-output (SISO) models~\cite{SOT, SOT4, SA-SOT, MT-whisper, domsot} adopt serialized output training (SOT) to implicitly handle speaker separation via attention, producing unified transcription sequences with speaker-delimiter tokens. This paradigm offers greater flexibility than SIMO, as it does not require predefined speaker counts and can naturally handle varying numbers of speakers. Recent works have significantly advanced SOT-based MTASR: MT-LLM~\cite{LLM-SOT} integrates large language models to understand multi-talker speech; SACTC~\cite{sactc} and SD-CTC~\cite{SDCTC} employ enhanced Connectionist Temporal Classification for explicit speech separation within the encoder; and CSE~\cite{cse-sot} incorporates SIMO principles by employing a dual-branch encoder for explicit feature disentanglement.

To further enhance performance in complex overlapping scenarios, explicit speaker modeling has proven essential~\cite{e2e-sa, SSA}. For example, Self-Speaker Adaptation (SSA)~\cite{SSA} leverages speaker activity signals obtained from an auxiliary diarization model~\cite{sortformer} to inject speaker-specific kernels into the encoder. However, its reliance on an external diarization system for speaker activity estimation, together with its multi-instance inference design, where one model instance is deployed per speaker, introduces additional computational overhead and limits overall efficiency.

In parallel, the Mixture-of-Experts (MoE) paradigm~\cite{MoE, moelora} offers a promising direction for handling input variability via conditional computation. MoE has shown strong performance in ASR~\cite{ASR-MOE2-3, ASR-MOE5-5, HDMoLE, AV-ASR-MOE}. Ideally, MoE is well-suited for MTASR as it enables the dynamic allocation of specialized experts to handle varying numbers of speakers and different degrees of overlap. However, the application of MoE to MTASR remains largely unexplored, and a direct application faces a critical challenge: standard MoE routers typically rely on local layer inputs. In deep network layers, speaker-specific acoustic characteristics, which are essential for distinguishing speaker identity, are often diluted. This dilution makes it difficult for local routers to assign experts based on the speaker.

To bridge this gap, we propose \textbf{G}lobal-\textbf{L}ocal \textbf{A}ware \textbf{D}ynamic Mixture-of-Experts (GLAD). Motivated by the fact that shallow acoustic features inherently preserve rich speaker cues, we introduce a global router that processes these shallow representations to capture speaker-aware context. Simultaneously, local routers model layer-specific information derived from intermediate encoder layers. A dynamic fusion module then adaptively combines these signals per frame, guiding expert selection with both global speaker cues and local acoustic details. We apply GLAD to the SOT paradigm, resulting in GLAD-SOT. To facilitate future research, our code and data is publicly available\footnote{https://github.com/NKU-HLT/GLAD}. Our main contributions are as follows:

\begin{figure*}[t]
  \centering
  \includegraphics[width=0.95\textwidth]{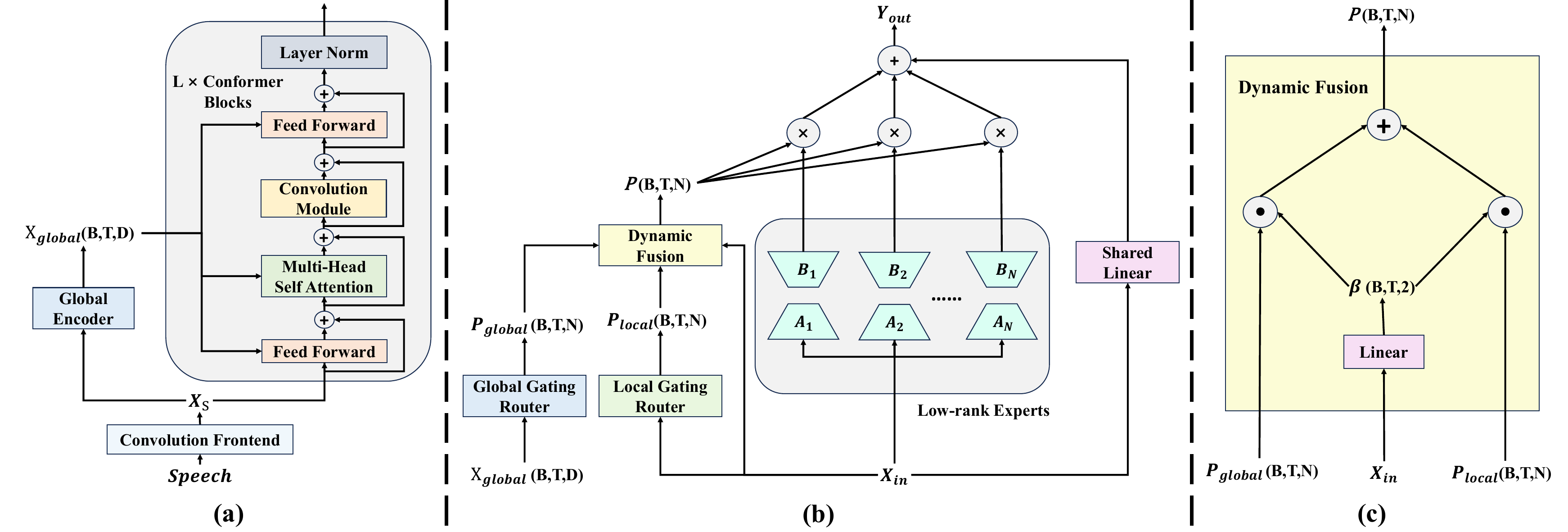}
  \caption{Overview of the proposed GLAD-SOT architecture. (a) A global linear encoder transforms features from the convolution frontend into a shared global representation, which is broadcast to each MoLE layer. (b) Each MoLE layer derives global weights from the shared global representation and integrates them with local signals to coordinate low-rank experts. (c) The global-local aware dynamic fusion module adaptively fuses weights to guide expert selection.}
  \label{fig:model}
  \vspace{-3mm}
\end{figure*}

\begin{itemize}
    \item To the best of our knowledge, this work represents the first application of MoE architectures in MTASR. Extensive experiments on LibriSpeechMix and CH109 demonstrate that our method outperforms strong SOT-based baselines, especially in challenging MTASR scenarios.
    \item We propose GLAD, a novel mechanism that dynamically fuses speaker-aware global context from shallow acoustic features with fine-grained local features. This dual-path routing strategy guides experts to disentangle overlapping speech by leveraging both speaker identity cues and phonetic details.
    \item We provide comprehensive ablation studies to validate the efficacy of our design. Our analysis reveals that incorporating global acoustic features is critical for speaker-aware expert routing, particularly in high-overlap scenarios where distinguishing speaker identity is most challenging.
\end{itemize}

\section{Our Method}
\label{sec:OurMethod}

\subsection{Serialized Output Training for Multi-talker ASR}
\label{subsec:sot}
Serialized Output Training (SOT)~\cite{SOT} offers a scalable solution for MTASR by using a single decoder to handle varying speaker counts, contrasting with conventional approaches that require multiple output branches. For multi-talker utterances, the SOT model is trained to predict a unified sequence concatenating all speakers' transcriptions, separated by a token $<sc>$, such as `$text1 <sc> text2 ...$'. In our experiments, the transcriptions follow the chronological order of speakers.

\subsection{Mixture of Low-rank Experts Integration in SOT}
\label{subsec:MoLE}
The Mixture-of-Experts (MoE) paradigm has proven effective in scaling neural networks by distributing computation across specialized modules. Building upon this concept, the Mixture of Low-Rank Experts (MoLE) framework substitutes traditional dense expert layers with low-rank decomposition modules. This design leverages the parameter-efficient structure of LoRA~\cite{LoRA} to enhance scalability while maintaining model capacity.

In our approach, we integrate MoLE with SOT by replacing all linear layers in the Conformer encoder~\cite{conformer} with MoLE modules. Each layer contains $N$ low-rank experts operating in parallel with a shared linear transformation. Given an input $X \in \mathbb{R}^{d_{in}}$, the MoLE layer output is formulated as:

\vspace{-1mm}
\begin{equation}
    Y_{out} = W_L X 
    + \frac{\alpha}{r}  {\textstyle \sum_{i=1}^{N}}  P_i B_i A_i X
    + b,
    \label{eq:mole}
\end{equation}
\vspace{-2mm}

\noindent where $W_L$ and $b$ denote the shared linear layer weight and bias. Each low-rank expert is parameterized by low-rank matrices $A_i \in \mathbb{R}^{r \times d_{in}}$ and $B_i \in \mathbb{R}^{d_{out} \times r}$ with rank $r \ll \text{min}(d_{in}, d_{out})$, scaled by a factor $\alpha$. The expert weight $P_i$ is computed via a gating mechanism combining global and local routing (see Sections~\ref{subsec:GR} and~\ref{subsec:GLFusion}).

\subsection{Gating Router Design}
\label{subsec:GR}

In MTASR, speaker-related information is critical for distinguishing overlapping speech. As shallow acoustic features contain rich speaker characteristics, we propose a global encoder that operates on the speech features $X_{S} \in \mathbb{R}^{T \times d_{h}}$ extracted by the convolution frontend. Unlike local routers that rely on intermediate representations, the global router captures speaker-aware context from the shallow features.

As shown in Fig.~\ref{fig:model}(a) and Fig.~\ref{fig:model}(b), we first employ a global linear encoder to map $X_{S}$ into the global representation $X_{global} = X_{S} W_{enc}$. This representation is then broadcast to each MoLE layer to independently compute the layer-specific global expert distribution $P_{global} \in \mathbb{R}^{T \times N}$:

\vspace{-1mm}
\begin{equation}
    P_{global} = \text{softmax}(\text{KeepTopK}(X_{global} W_{global}, K)),
    \label{eq:gr}
\end{equation}
\vspace{-2mm}

\noindent where $W_{enc}$ and $W_{global} \in \mathbb{R}^{d_h \times N}$ are weight matrices for the global encoder and router, respectively, and $N$ is the expert count. The $\text{KeepTopK}(\cdot, K)$ function preserves only the top $K$ values while setting others to $-\infty$. This enables the router to derive speaker-aware weights from shared features, providing guidance grounded in global context.

While the global router leverages the global representation to capture speaker context, the local router focuses on fine-grained acoustic patterns. Specifically, it computes local expert probabilities $P_{local} \in \mathbb{R}^{T \times N}$ from the intermediate input $X_{in}$:

\vspace{-1mm}
\begin{equation}
    P_{local} = \text{softmax}(\text{KeepTopK}(X_{in} W_{local}, K)),
    \label{eq:lr}
\end{equation}
\vspace{-2mm}

\noindent where $W_{local} \in \mathbb{R}^{d_h \times N}$ is the weight matrix of the local router.

\subsection{Global-Local Aware Dynamic Fusion}
\label{subsec:GLFusion}

To harness the complementary strengths of the speaker-aware global context and fine-grained local features, we introduce a Global-Local Aware Dynamic Fusion mechanism. It dynamically balances the routing signals from both streams at each frame, guiding expert activation with a fused perspective.

As shown in Fig.~\ref{fig:model}(c), we employ a dynamic fusion module to generate gating weights based on the local input $X_{in}$:

\vspace{-1mm}
\begin{equation}
    \beta = \text{softmax}(X_{in} W_{fusion}),
    \label{eq:fusionweight}
\end{equation}
\vspace{-2mm}

\noindent where $W_{fusion} \in \mathbb{R}^{d_{h} \times 2}$ is a trainable fusion matrix. The output $\beta \in \mathbb{R}^{T \times 2}$ is decomposed into $\beta_{g}$ and $\beta_{l}$, representing the dynamic fusion weights for global and local distributions, respectively. The final probability for the $i$-th expert, denoted as $P_{i}$, is then computed via weighted summation:

\vspace{-1mm}
\begin{equation}
    P_{i} = \beta_{g} \odot P_{global,i} + \beta_{l} \odot P_{local,i},
    \label{eq:gl_fusion}
\end{equation}
\vspace{-2mm}

\noindent where $\odot$ denotes element-wise multiplication. This dynamic fusion allows each expert to adaptively leverage global speaker cues and local acoustic details for more effective routing.

\subsection{Training Loss}
To prevent routing collapse and ensure expert specialization, we incorporate the auxiliary load-balancing loss as proposed in Switch Transformer~\cite{switchtransformer}:

\vspace{-1mm}
\begin{equation}
    \mathcal{L}_{bal} = \gamma N \sum_{i=1}^{N} f_i \overline{P_i} ,
    \label{eq:balance_loss}
\end{equation}
\vspace{-2mm}

\noindent where $f_i$ denotes the fraction of tokens routed to expert $i$ via the top-$k$ mechanism, and $\overline{P_i}$ represents the mean routing probability for that expert across the batch. In our experiments, the balancing weight $\gamma$ is set to 0.01. 

The total training loss is then formulated as the sum of the ASR loss and the load-balancing term:

\vspace{-1mm}
\begin{equation}
    \mathcal{L} = \mathcal{L}_{ASR} + \mathcal{L}_{bal}
    \label{eq:finalloss}
\end{equation}
\vspace{-2mm}

\section{Experimental Setup}
\label{sec:ExperimentalSetup}

\subsection{Dataset}
\label{subsec:dataset}

We conduct experiments on LibriSpeechMix (LSM)~\cite{SOT}, a standard MTASR benchmark derived from LibriSpeech~\cite{Librispeech} via simulated overlapping speech. Following prior work~\cite{SOT,cse-sot,sactc}, two-speaker mixtures are created by randomly pairing training utterances with random temporal offsets. The final training set combines randomly sampled subsets of the two-speaker mixtures and original single-speaker utterances, totaling 1.35k hours (0.69k hours 1-speaker and 0.66k hours 2-speaker).
For evaluation, we follow~\cite{cse-sot,sactc} and group test samples by overlap ratio (overlap duration over total length) into low $(0,0.2]$, medium $(0.2,0.5]$, and high $(0.5,1.0]$ categories, enabling systematic robustness analysis under varying overlap conditions.

For evaluation on real-world data, we use CH109, a two-speaker subset comprising 109 sessions from the 120-session Callhome American English Speech (CHAES) corpus\footnote{\url{https://catalog.ldc.upenn.edu/LDC97S42}}. To facilitate model adaptation to real-world acoustic and conversational conditions, we construct the CH11-mix dataset using the remaining 11 CHAES sessions.

\begin{table*}[t]
\captionsetup{aboveskip=2pt, belowskip=2pt}  
\caption{WER(\%) Results on Librispeech, LibrispeechMix-2mix and LibrispeechMix-3mix. All models are trained on single-talker and two-talker data. The best results are highlighted in bold and the second-best results are indicated with an underline.}
\label{tab:res}
\renewcommand{\arraystretch}{1.35}
\centering
\resizebox{\textwidth}{!}{%
\begin{tabular}{c|l|c|cc|cccccc|cccccc}
\hline
\multirow{3}{*}{System} & \multicolumn{1}{c|}{\multirow{3}{*}{Method}} & \multirow{3}{*}{Parm.(M)} & \multicolumn{2}{c|}{Librispeech}             & \multicolumn{6}{c|}{LibrispeechMix-2mix}                                                                     & \multicolumn{6}{c}{LibrispeechMix-3mix (\textbf{Generalization})}                                                           \\ \cline{4-17} 
                        & \multicolumn{1}{c|}{}                        &                           & \multirow{2}{*}{Dev} & \multirow{2}{*}{Test} & \multicolumn{2}{c|}{Overall}                     & \multicolumn{4}{c|}{Test(Conditional)}                    & \multicolumn{2}{c|}{Overall}                       & \multicolumn{4}{c}{Test(Conditional)}                         \\ \cline{6-17} 
                        & \multicolumn{1}{c|}{}                        &                           &                      &                       & Dev          & \multicolumn{1}{c|}{Test}         & low          & mid          & high         & OA-WER       & Dev           & \multicolumn{1}{c|}{Test}          & low           & mid           & high          & OA-WER        \\ \hline
S1                      & SOT~\cite{SOT}                                          & 36.07                     & 5.3                  & 6.2                   & 7.9          & \multicolumn{1}{c|}{8.1}          & 8.0          & 7.2          & 10.7         & 8.6          & 23.9          & \multicolumn{1}{c|}{23.5}          & 23.6          & 21.9          & 26.9          & 24.1          \\
S2                      & CSE-SOT~\cite{cse-sot}                                      & 38.90                     & 4.3                  & {4.5}             & {\underline{7.5}}    & \multicolumn{1}{c|}{{\underline{7.3}}}    & {\underline{6.3}}    & {7.3}    & 10.6         & {\underline{8.1}}    & 23.8          & \multicolumn{1}{c|}{24.2}          & \textbf{18.2}          & 23.6          & 31.5          & 24.4          \\
S3                      & SOT+SACTC-12~\cite{sactc}                                    & 34.18                     & 4.1                  & 4.4                   & 8.0          & \multicolumn{1}{c|}{7.8}          & 6.6          & 7.6          & 11.6         & 8.6          & {22.8}    & \multicolumn{1}{c|}{22.6}          & {\underline{18.7}}    & {21.9}    & 28.2          & 22.9          \\
S4                      & SOT+SACTC-13~\cite{sactc}                                    & 35.77                     & 4.1                  & \underline{4.3}                   & 7.8          & \multicolumn{1}{c|}{7.7}          & 6.7          & 7.6          & 10.5         & 8.3          & {\underline{21.9}}    & \multicolumn{1}{c|}{22.2}          & {19.1}    & {\underline{21.0}}    & 27.8          & 22.6          \\
S5                      & GLAD-SOT                                     & 35.31                     & \textbf{3.7}         & \textbf{4.1}          & \textbf{7.2} & \multicolumn{1}{c|}{\textbf{6.8}} & \textbf{5.7} & \textbf{6.9} & \textbf{9.7} & \textbf{7.4} & \textbf{21.7} & \multicolumn{1}{c|}{\textbf{21.1}} & \textbf{18.2} & \textbf{20.0} & \textbf{26.3} & \textbf{21.5} \\ \hline
S6                      & \quad w/o global router       & 35.09                     & 4.1                  & 4.5                   & 8.3          & \multicolumn{1}{c|}{8.7}          & 8.4          & 7.8          & 11.6         & 9.3          & 24.6          & \multicolumn{1}{c|}{24.5}          & 21.8          & 23.6          & 29.1          & 24.8          \\
S7                      & \quad w/o dynamic fusion      & 35.23                     & {\underline{3.9}}            & \underline{4.3}                   & \textbf{7.2} & \multicolumn{1}{c|}{7.5}          & 6.8          & {\underline{7.1}}    & {\underline{10.3}}   & {\underline{8.1}}    & \textbf{21.7} & \multicolumn{1}{c|}{{\underline{22.0}}}    & 19.7          & {21.1}    & {\underline{26.4}}    & {\underline{22.4}}    \\
S8                      & \quad w GLAD in FFN            & 34.63                     & 4.3                          & 4.6                          & 8.5                          & \multicolumn{1}{c|}{10.0}                         & 9.1                          & 9.1                          & 14.9                          & 11.0                         & 23.1                          & \multicolumn{1}{c|}{24.5}                          & 23.5                          & 23.8                          & 27.1                          & 24.8                          \\
S9                      & \quad w GLAD in Att            & 33.65                     & 4.2                          & 4.5                          & 9.6                          & \multicolumn{1}{c|}{11.2}                         & 12.6                         & 9.6                          & 11.1                          & 11.1                         & 24.4                          & \multicolumn{1}{c|}{24.2}                          & 21.4                          & 23.3                          & 28.9                          & 24.5                          \\ \hline
\end{tabular}%
}
\vspace{-2mm}
\end{table*}

\subsection{Model Settings}
\label{subsec:modelsettings}
We implement our Conformer-based models using ESPnet2~\cite{espnet}. The base architecture comprises a Conformer encoder and a 6-block Transformer decoder, both utilizing 4-head self-attention with 256 hidden units. The Macaron-style encoder blocks have an FFN dimension of 1024, while the decoder uses 2048. In our 12-block GLAD-SOT, all encoder linear layers are replaced by MoLE modules containing 3 experts ($r=8, \alpha=8$).
To ensure fair evaluation under comparable parameter constraints, we reproduce several strong Conformer-based baselines using their original optimal hyperparameters: SOT (14 encoder blocks), CSE-SOT (13 blocks), and SOT+SACTC (12- and 13-block variants).

\subsection{Training Settings and Metrics}
\label{subsec:trainingsettingsandmetics}

Training Setup: All models are trained for 50 epochs in bfloat16 mixed precision on 8 NVIDIA GeForce RTX 3090 GPUs using the Adam optimizer ($lr=5e^{-4}$, 25k warm-up steps). The final models are obtained by averaging the top-10 checkpoints selected based on validation performance on dev-clean-2mix set. 

Evaluation Metrics: We report standard Word Error Rate (WER) for single-talker tasks. For multi-talker scenarios, we adopt Permutation-Invariant WER (PI-WER)~\cite{SOT}. However, given the uneven distribution of overlap levels in the test set, PI-WER may not fully reflect robustness. Therefore, following~\cite{cse-sot,sactc}, we utilize Overlap-Aware WER (OA-WER) to provide a balanced evaluation across varying overlap conditions.

Experimental Design: In our primary experiments, models are trained on single- and two-speaker data. We evaluate them on LSM-1/2mix and assess generalization on LSM-3mix.

\section{Results and Discussion}
\label{sec:ResultsAndDiscussions}

\subsection{Performance of GLAD-SOT}
In Table~\ref{tab:res} (S1–S5), we compare GLAD-SOT with several strong SOT-based baselines. GLAD-SOT achieves the best overall performance across all evaluated scenarios.

In the LSM-2mix setting, CSE-SOT (S2) achieves competitive performance due to its dual-branch architecture that explicitly separates speakers. However, its two-branch design limits scalability, leading to performance degradation in LSM-3mix-mid/high scenarios. In contrast, GLAD-SOT (S5) not only further improves performance in two-speaker settings but also maintains strong results in three-speaker scenarios. This suggests that leveraging global contextual information to guide dynamic expert routing provides greater flexibility than explicit branch separation.

In LSM-3mix evaluations, SOT+SACTC (S3, S4) demonstrates improved generalization by encouraging the encoder to disentangle overlapping speech into single-speaker representations. Nevertheless, GLAD-SOT consistently outperforms these methods, particularly under high-overlap conditions. Its superior performance on LSM-3mix indicates that the proposed GLAD framework effectively enhances robustness and generalization in complex MTASR scenarios.

\subsection{Ablation Study}
\label{subsec:Ablation Study}

\textbf{Impact of global information}: To investigate the role of global routing, we compare S5 and S6. In S6, we ablate the global encoder and remove the global context from the MoLE modules, relying solely on local information. As shown in Table~\ref{tab:res}, S5 significantly outperforms S6, confirming the necessity of the global context. This improvement stems from the global router's ability to capture speaker-aware characteristics from the speech signal and direct experts to focus on distinct speakers.

\textbf{Effect of Global-Local Aware Dynamic Fusion}: To evaluate the necessity of the dynamic fusion module, we compare configurations S5 and S7. In S7, the final expert probabilities are computed through an unweighted static summation of the global and local weights ($P = P_{local} + P_{global}$). This fixed combination yields inferior performance, particularly on the LSM-2/3mix-low set, due to its inability to adaptively balance the two information streams. Specifically, while fine-grained local acoustic details are crucial for low-overlap scenarios, global speaker context becomes more critical as the overlap ratio increases. In contrast, the dynamic fusion mechanism in S5 adaptively adjusts the routing weights based on both streams, achieving robust performance across diverse overlap conditions.

\textbf{Impact of GLAD placement}: We compare S5, S8, and S9 to study where GLAD is most effective. S8 and S9 apply GLAD to FFN and attention modules, respectively, while S5 applies it to both. 
As Table~\ref{tab:res} shows, S8 outperforms S9 in LSM-2mix-low/mid scenarios, indicating the FFN's strength in feature transformation. Conversely, S9 performs better in LSM-2mix-high and LSM-3mix-low/mid, underscoring the role of attention in disentangling overlapping speech. However, S9's advantage diminishes in the challenging zero-shot 3-mix-high scenario, suggesting that attention alone is insufficient under extreme acoustic complexity. Finally, S5 achieves the lowest WER, confirming that the synergy between attention-driven disentanglement and FFN-based transformation provides the most robust and complementary improvements.

\subsection{Performance in Real World}
\begin{table}[t]
\captionsetup{aboveskip=2pt, belowskip=2pt}
\caption{WER (\%) results on the CH109. The symbol $*$ indicates that the model is fine-tuned on the CH11-mix dataset for 5 epochs. The best results are highlighted in bold.}
\label{tab:realworld}
\centering
\renewcommand{\arraystretch}{1.05}
\resizebox{0.45\textwidth}{!}
{%
\begin{tabular}{l|cccc}
\hline
\multicolumn{1}{c|}{\multirow{2}{*}{Method}} & \multicolumn{4}{c}{CH109}               \\ \cline{2-5} 
\multicolumn{1}{c|}{}                        & 1-speaker & 2-speaker & Total & Average \\ \hline
SOT$^{*}$~\cite{SOT}                                    & 40.8      & 52.7      & 45.4  & 46.8    \\
CSE-SOT$^{*}$~\cite{cse-sot}                                & 36.9      & 50.8      & 40.0  & 43.9    \\
SOT+SACTC-13$^{*}$~\cite{sactc}                              & 33.0      & 50.7      & 37.4  & 41.9    \\
GLAD-SOT$^{*}$                               & \textbf{32.5}      & \textbf{48.9}      & \textbf{36.6}  & \textbf{40.7}    \\ \hline
\end{tabular}%
}
\vspace{-2mm}
\end{table}

To assess the model's performance in real-world scenarios, we fine-tune the SOT~\cite{SOT}, CSE-SOT~\cite{cse-sot}, SACTC-13~\cite{sactc}, and our GLAD-SOT models (from Table~\ref{tab:res}) on the CH11-mix dataset for 5 epochs and evaluate them on the CH109 set.

As shown in Table~\ref{tab:realworld}, GLAD-SOT achieves the best performance in both 1-speaker and 2-speaker scenarios. Notably, in the challenging 2-speaker overlapping condition, GLAD-SOT significantly outperforms all baseline models, demonstrating its strong robustness and great potential for practical deployment.

\subsection{Visual Analysis}

To investigate the respective contributions of global and local information in expert selection, we visualize the global fusion weight ($\beta_g$) from Eq.~\ref{eq:gl_fusion} at encoder layers 1, 8, and 12 under varying overlap conditions on LSM-test-2mix and 3mix. 
Specifically, we analyze the $\beta_g$ values from the MoLE modules substituted for the attention output projection layers ($W_O$), as~\ref{subsec:Ablation Study} shows that attention mechanism is cruial for separate overlap speech. To obtain a representative metric, we average the frame-level $\beta_g$ values across the test sets using mean pooling. Given the constraint that $\beta_g + \beta_l = 1$, we focus our discussion on $\beta_g$ to characterize the dynamic balance between global and local streams.

\begin{figure}[t]
    \centering
    \begin{subfigure}[b]{0.49\linewidth}
        \centering
        \includegraphics[width=\linewidth]{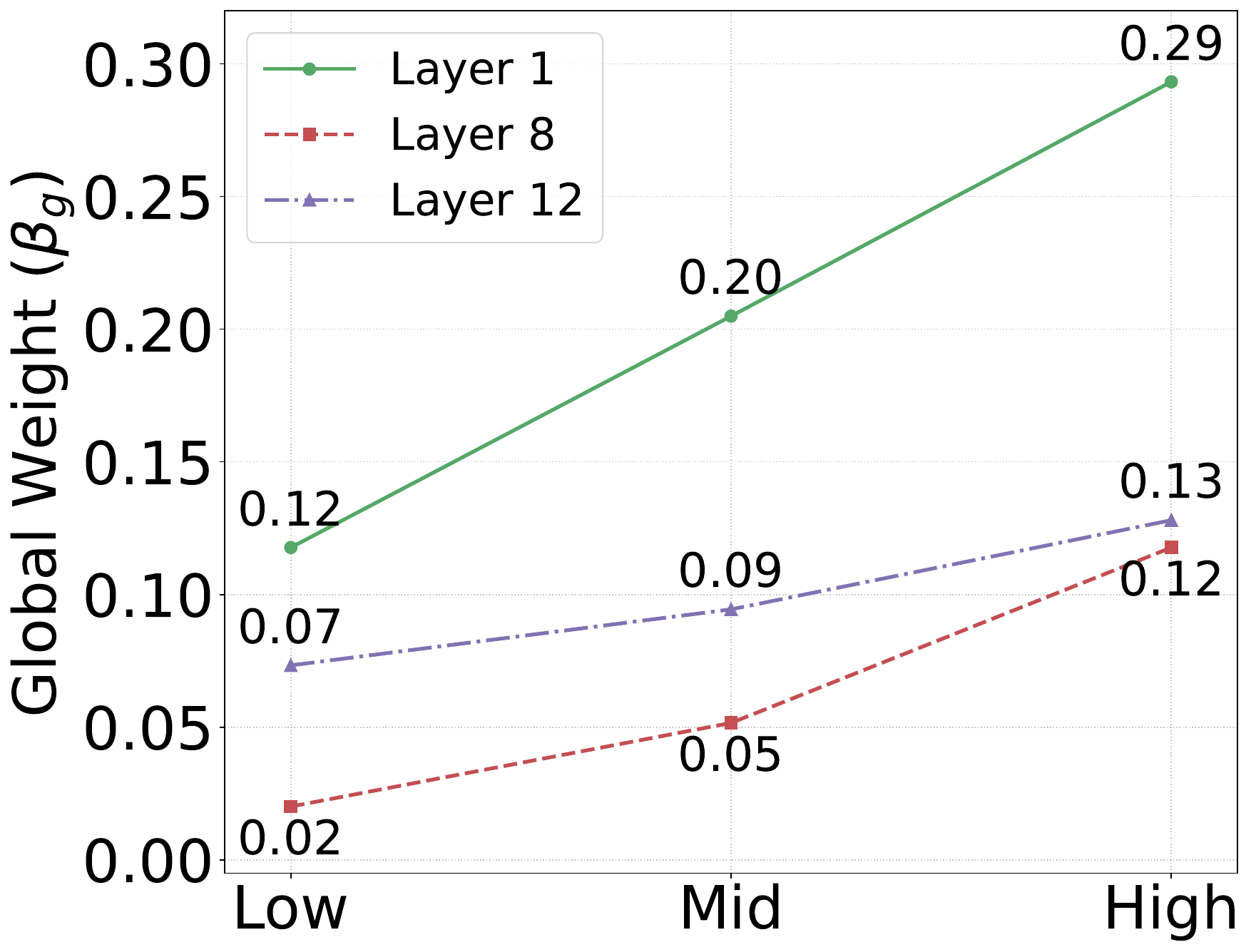}
        \caption{LSM-test-2mix}
        \label{fig:left}
    \end{subfigure}%
    \hfill
    \begin{subfigure}[b]{0.49\linewidth}
        \centering
        \includegraphics[width=\linewidth]{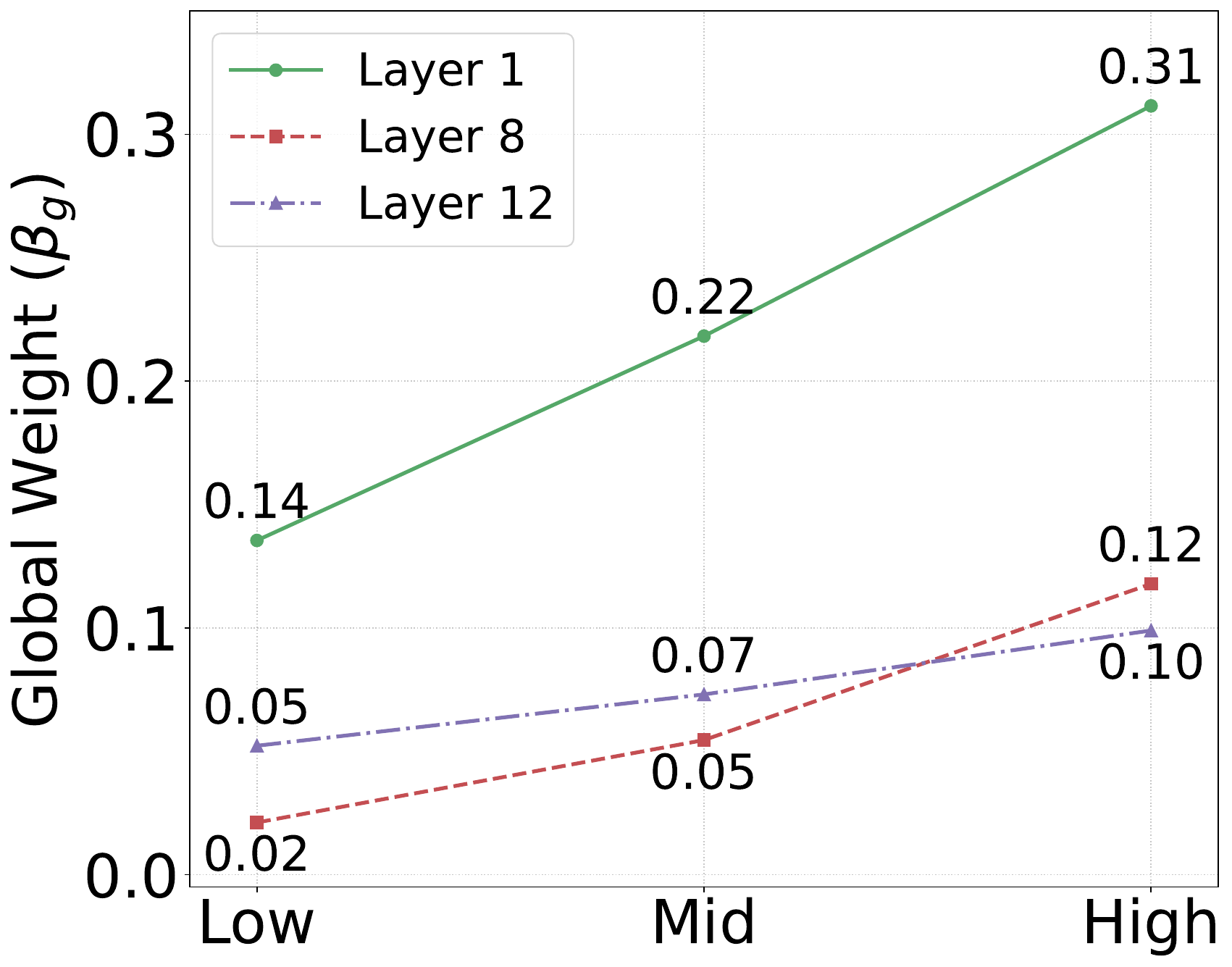}
        \caption{LSM-test-3mix}
        \label{fig:right}
    \end{subfigure}
    \captionsetup{aboveskip=2pt, belowskip=2pt}
    \caption{Visualization of the average global fusion weight ($\beta_g$) at different encoder layers across varying overlap scenarios.}
    \label{fig:lab-visual}
    \vspace{-3mm}
\end{figure}

Fig.~\ref{fig:lab-visual} demonstrates consistent trends across both LSM-2mix and LSM-3mix. First, $\beta_g$ increases with the overlap ratio, supporting our hypothesis that global speaker-specific features are critical for distinguishing identities in highly mixed audio. Second, the layer-wise distribution of $\beta_g$ reveals a nuanced pattern of information utilization. At Layer 1, $\beta_g$ reaches its highest value, indicating that shallow layers rely heavily on global speaker cues to initiate separation from fully mixed signals. At the intermediate Layer 8, $\beta_g$ decreases, suggesting that partial disentanglement by earlier layers reduces the immediate reliance on global anchors. Interestingly, at Layer 12, $\beta_g$ increases again compared to Layer 8. This rebound implies that speaker-specific information may become attenuated in deeper layers, requiring renewed global context to preserve speaker identity and stabilize expert routing.

\section{Conclusions}
\label{sec:conclusions}
In this paper, we propose GLAD, a novel approach that leverages Mixture of low-rank Experts to address the multi-talker ASR task by dynamically integrating global and local information. Extensive experiments on LibriSpeechMix and CH109 demonstrate that our GLAD mechanism significantly outperforms existing SOT-based MTASR methods, especially in challenging multi-talker scenarios.

\section{Generative AI Use Disclosure}
During the preparation of this manuscript, the authors used generative AI tools solely for language editing and readability improvement. All generated suggestions were carefully reviewed and revised by the authors. The authors take full responsibility for the accuracy, integrity, and originality of the work. No generative AI was used to generate any substantive scientific content or significant portions of the manuscript.

\bibliographystyle{IEEEtran}
\bibliography{mybib}

\end{document}